%% file: teaching_paper.tex
\documentclass[copyright,creativecommons]{eptcs}
\providecommand{\event}{SOS 2021} 

\usepackage{breakurl}             
\usepackage{underscore}           

\usepackage{booktabs}   
\usepackage{subcaption} 

\usepackage{mathtools} 
\usepackage{semantic}
\usepackage{stmaryrd}
\usepackage{amsmath}
\usepackage{amssymb}

\usepackage{color}
\usepackage{listings}
\usepackage{mathpartir}

\usepackage{pifont}

\usepackage{tcolorbox}
\usepackage{tabularx}
\usepackage{array}
\usepackage{colortbl}
\tcbuselibrary{skins}

\usepackage{longtable}
\usepackage{clrscode3e}

\input{listing-macros.tex}

\input{definitions.tex}

\title{Language Transformations in the Classroom}
\author{Matteo Cimini
\institute{University of Massachusetts Lowell\\ Lowell, Massachusetts, USA}
\email{matteo\_cimini@uml.edu}
\and
Benjamin Mourad
\institute{University of Massachusetts Lowell\\ Lowell, Massachusetts, USA}
\email{benjamin\_mourad@student.uml.edu}
}
\def\titlerunning{Language Transformations in the Classroom}
\def\authorrunning{M. Cimini \& B. Mourad}

\begin{document}
\maketitle

\begin{abstract}
Language transformations are algorithms that take a language specification in input, and return the language specification modified. 
Language transformations are useful for automatically adding features such as subtyping to programming languages (PLs), and for automatically deriving abstract machines. 

In this paper, we set forth the thesis that teaching programming languages features with the help of language transformations, in addition to the planned material, can be beneficial for students to help them deepen their understanding of the features being taught.

We have conducted a study on integrating language transformations into an undergraduate PL course. We describe our study, the material that we have taught, and the exam submitted to students, and we present the results from this study.  
Although we refrain from drawing general conclusions on the effectiveness of language transformations, this paper offers encouraging data. We also offer this paper to inspire similar studies. 

\end{abstract}

\section{Introduction}

Computer Science university curricula include undergraduate courses in programming languages (PLs). These courses vary greatly in the content they offer, and they may also have various names such as ``Principles of Programming Languages'', and ``Organization of Programming Languages'', to make some examples. 
Typically, the goal of these courses is not to teach one specific PL. Conversely, students are exposed to the conceptual building blocks from which languages are assembled, the various programming paradigms that exist, and students are challenged to think about various PL features in their generality. 

It is typical for these courses to cover PL features such as subtyping, abstract machines, type inference, parametric polymorphism, as well as many others. 
Some of these features can be regarded as variations on a base PL. For example, it is not uncommon to design a PL, and add subtyping \emph{afterwards}. 
It is then interesting to understand what are the modifications that need to take place in order to incorporate subtyping in that base language. 
A good way to analyze this is by looking at how formal typing rules need to change. 
Consider, for example, the typing rule of function application below on the left, and its version with (algorithmic) subtyping on the right. 

{ 
\begin{gather*}
{
\ninference{t-app}
	{
	\Gamma \typeOf \app e_1 : T_1\to T_2 \\\\
	\Gamma \typeOf \app e_2 : T_1 	
	} 
	{ \Gamma \typeOf \app e_1\app e_2 : T_2}
}
\quad \Longrightarrow \quad 
{
\ninference{t-app'}
	{
	\Gamma \typeOf \app e_1 : T_{11}\to T_2 \\\\
	\Gamma \typeOf \app e_2 : T_{12} \\  T_{12} <: T_{11}
	} 
	{ \Gamma \typeOf \app e_1\app e_2 : T_2}
}
\end{gather*}
}

\textsc{(t-app)} rejects programs that pass an integer to a function that works on floating points, such as the program $((\lambda x:\key{float}.x) \app 3)$, where $\emptyset \typeOf 3 : \key{int}$. 
This is because the type $T_1$ in $T_1 \to T_2$, which is the domain of the function $e_1$, needs to be the exact same type $T_1$ of the argument $e_2$. 
If we were to add subtyping, such a parameter passing would be accepted by the type system. 

The first modification that \textsc{(t-app')} makes of \textsc{(t-app)} is to let the domain of the function and the argument have different types. 
To do so, \textsc{(t-app')} assigns two different variables to the two occurrences of $T_1$, that is, $T_{11}$ for the domain of the function, and $T_{12}$ for the type of the argument. 
Next, \textsc{(t-app')} needs to understand how $T_{11}$ and $T_{12}$ are related by subtyping. As $T_{11}$ appears in contravariant position in $T_{11}\to T_2$, 
it means that $T_{11}$ describes the type of an input. The argument $e_2$ will be provided as a value. Therefore, it is the type $T_{12}$ of the argument that must be a subtype of $T_{11}$, rather than the other way around, for example. Hence, the subtyping premise $T_{12} <: T_{11}$ is added to \textsc{(t-app')}. 
 
We can describe these modifications with an algorithm that takes \textsc{(t-app)}, and automatically transforms it into \textsc{(t-app')}. To summarize, such an algorithm must perform two steps: 

\begin{itemize}
\item \textbf{Step 1}: Split equal types into fresh,  distinct variables, and 
\item \textbf{Step 2}: Relate these new variables by subtyping according to the variance of types. 
\end{itemize}


This type of algorithm can be formulated over a formal data type for language specifications. In other words, we can devise a procedure that takes a language specification in input (as a data type), and returns another language specification (with subtyping added). 
These algorithms are called \emph{language transformations} \cite{MouradC20}. 
One of the benefits of language transformations is that they do not apply just to one language. Instead, they can apply to several languages. 
For example, the two steps above can add subtyping for types other than the function type, such as pairs, lists, option types, and other simple types. 

\paragraph{Our Thesis}  
Another benefit of language transformations is that they highlight the central insights behind a feature being added. 
For example, \textbf{Step 1} and \textbf{Step 2} are key aspects of subtyping. Teaching students the algorithms that automatically apply \textbf{Step 1} and \textbf{Step 2} to languages can provide them with a firmer grasp of the concept of subtyping overall. 

The approach is not limited to subtyping. The adding of other PL features can be formulated as language transformations, and taught to students in class as well. 
We think that exposing students to the language transformations for adding PL features may constitute a good addition in the classroom. 

In this regard, however, we point out that we do not advocate for teaching PL features exclusively with the {sole} help of language transformations. 
For example, we teach subtyping using the material in the TAPL textbook \cite{tapl}, and we are skeptical that it would be a good idea to skip this material before introducing language transformations. 
This is because language transformations constitute quite a technical deep dive, and students could benefit from a gentler introduction of PL concepts. 

Ultimately, in this paper we set forth the thesis that \emph{using language transformations for teaching PL features, in addition to the planned material, can be beneficial for students to deepen their understanding of the features being taught.} 

\paragraph{Contributions of this Paper}

We have experimented with teaching the language transformations for adding subtyping and deriving CK abstract machines \cite{FelleisenFlatt02}. 
We have conducted our study on two instances of an undergraduate course on programming languages. 

In class, we first have introduced subtyping with material from TAPL \cite{tapl}, as mentioned above, and then we have taught the language transformations for adding subtyping (which we describe in Section 2.1). To evaluate whether our students gained a good understanding of subtyping, the final exam presented them with a language with operators that are not standard. Then, the exam asked students to add subtyping to these operators based on the language transformations that they have learned. 

In the context of this study, we have collected information about students' success in providing a correct answer to such a task.  
We describe the final exam in detail in Section 2.3, and we report on the results of this study in Section 3. 

We have taught the topic of CK machines following the notes of Felleisen and Flatt \cite{FelleisenFlatt02}. We then have taught the language transformations for deriving CK machines (which we describe in Section 2.2). Analogously to subtyping, the final exam asked our students to derive the CK machine for a language with operators that are not standard. We then have collected information about students' exam answers for this task, and we report on this data in Section 3. In total, the study involved 55 undergraduate students. 

To summarize our contributions, in this paper: 

\begin{itemize}
\item We set forth the thesis that language transformations can be a beneficial addition in PL courses, as formulated above. 
\item We describe the study that we have conducted, which includes the material that we have taught, and the exam submitted to the students. 
\item We present the results from our study. Although we explicitly say that we should not consider our results conclusive, the data that we present is encouraging. 
\item We offer this paper to inspire similar studies towards gathering evidence for, or against, our thesis.

\end{itemize}

\paragraph{Roadmap of the Paper}
Section 2 describes the study that we have conducted, Section 3 presents our results, Section 4 describes our future work, and Section 5 concludes the paper. 

\leave{
\begin{itemize}
\item courses in PL teach features: subtyping etc. 
\item some features (in the collogquial sense). we need to be careful | can be described as a variant of a language. 
\item example: subtyping
\item algo intuitive.
\item benefits of lang transformations: apply to many, bring up insights. 
\item THESIS: lang transformation => beneficial in teaching!
\item 1 course experiment, preliminary
\item no data.. didnt fo welll. COVID??  however, we share. 
\item GOALS: set forth the thesis. tell what we are doing. open the conversation. inspire instructors. 
\item in the rest of the paper: WHAT we have done. OUR PLAN for future. DIFFERENT IDEAS. 
\end{itemize}
}

\section{Language Transformations in Class}

\paragraph{General Details about the Course}
The course is at the undergraduate level, and is based on the TAPL textbook \cite{tapl}. The course covers the typical topics of PL theory on defining syntax (BNF grammars), operational semantics, and type systems of PLs. 
The course also covers several other topics such as parameter passing, scoping mechanisms, subtyping, abstract machines, recursion, exceptions, dynamic typing, memory management, concurrency, and logic programming. 
Students are then familiar with the formalisms of operational semantics and type systems when the course covers the topics of subtyping and abstract machines.

The evaluations of the course include a long-term programming project in which students develop an interpreter for a functional language with references in OCaml, and a final exam with questions and open answers at the end of the course.  
The final exam tests our students on the topics of subtyping and abstract machines. We will describe our exam in Section 2.3.

\paragraph{Algorithms in Pseudo-Code}

Language transformations are algorithms, which begs the question on what syntax we should use to describe them. 
We took inspiration from courses in Algorithms and Data Structures, and from textbooks such as \cite{clrs}, 
where algorithms are described in pseudo-code. Therefore, we have used a pseudo-code that, to our estimation, 
was always intuitive to students, even though we did not thoroughly and precisely define it (as in \cite{clrs}). 

\paragraph{Language Specifications}

During the course, students acquire familiarity with formal definitions of programming languages, which they learn through TAPL. 
To recap, languages are defined with a BNF grammar and a set of inference rules. Inference rules are used to define a type system, a reduction relation, and auxiliary relations, if any. 
To make an example, we repeat the typical formulation of the simply-typed $\lambda$-calculus. 
We use a small-step operational semantics and evaluation contexts. \\
(Below, $B$ is some base type.)

{
\begin{syntax}
  \text{Type} & T & ::= & T \to T \mid B  \\
  \text{Expression} & e & ::= & x  \mid \lam{x\of T} e \mid e~e \\
  \text{Value} & v & ::= & \lam{x\of T} e \\
  \text{Evaluation Context} & E & ::= & [ \cdot ]  \mid E~e\mid v~E 
\\
  \text{Type Environment} & \Gamma & ::= & \emptyset \mid \Gamma, x:T 
\end{syntax}
\begin{gather*}
\inference{x:T \in \Gamma}{\Gamma\type x:T}
\quad
\inference{\Gamma, x:T_1\type  e : T_2}
          {\Gamma\type (\lam{x\of T_1} e) : T_1 \to T_2}
\quad
\ninference{t-app}{\Gamma \type e_1: T_{11} \to T_{12} \\
           \Gamma \type e_2: T_{11}} 
          {\Gamma \type e_1 \app e_2 : T_{12}}
\\[2ex]
((\lambda x\of T.e)~v) \step e[v/x]
\quad
\inference{e \step e'}
          {E[e] \step E[e']}
\end{gather*}
}

We allow our pseudo-code to refer to parts of a language specification. For example, if the language $L$ is the formulation of the simply-typed $\lambda$-calculus, then it contains both the grammar and the set of inference rules above. 
$L.rules$ retrieves the set of inference rules. Given a rule $r$, say \textsc{({t-app})}, $r.premises$ retrieves the set of premises of \textsc{({t-app})}, which are the formulae above the horizontal line. $r.conclusion$ retrieves the formula below the horizontal line. 
To our estimation, these references in the pseudo-code, as well as all other references, are rather intuitive, and they will be clear when we use them. (This is also the take on pseudo-code that \cite{clrs} has, where a number of operations are not defined beforehand.) 

\paragraph{Roadmap of this Section}
Below, we describe our experiment on teaching subtyping (Section \ref{subtyping}) and CK machines (Section \ref{ck}). We also describe the final exam given to students (Section \ref{exam}). 
Our pseudo-code for adding subtyping is based on an algorithm expressed in a domain-specific language in \cite{MouradC20}. 
We are not aware of any analogous algorithm that corresponds to our pseudo-code for deriving CK machines.  
The next sections describe the algorithms that we have taught in class, of which we do not claim any theoretical results of correctness.

\subsection{Language Transformation for Subtyping}\label{subtyping}

In class, we have taught subtyping based on the corresponding chapters in the TAPL textbook \cite{tapl}. 
Then, we have taught language transformation algorithms for adding subtyping to simple functional languages.  
The task of adding subtyping has been divided into the two steps that we have discussed in the introduction: 1) Split equal types, and 2) Relate new variables by subtyping according to the variance of types. 

\paragraph{Split Equal Types}

This step modifies the typing rules of a language so that the variables that occur more than once in their premises are given different variable names. 
As we have seen in the introduction, this is the first step to let different expressions have different types. 
We define the procedure $\proc{Split-Equal-Types}$ to perform this step. 
The pseudo-code of $\proc{Split-Equal-Types}$ is given below, which we explain subsequently. 
{
\begin{codebox}
  \Procname{$\proc{Split-Equal-Types}(P)$}
  \li $\textit{newPremises} \gets \emptyset, \textit{varmap} \gets \emptyset$
  \li \For $\text{each } p \in P$ \Do
  \li \If $p \text{ is of the form } \Gamma \vdash e : \textit{someType}$ \Do
  \li \For $\text{each } T \in \textit{someType} \text{ s.t. $T$ appears more than once in $P$} $ \Do 
  \li $T' \gets \proc{Fresh}(P)$
  \li $p \gets p \text{ where } T' \text{ replaces } T \text{ in } \textit{someType}$
  \li $\textit{varmap}(T) \gets \textit{varmap}(T) \cup \{T'\}$
  \End
  \End
  \li $\textit{newPremises} \gets \textit{newPremises} \cup \{p\}$
  \End
  \End
  \li \Return $(\textit{newPremises},\textit{varmap})$
\end{codebox}
}

$\proc{Split-Equal-Types}$ takes a set of premises $P$ in input, and returns a pair with two components: a set of premises \textit{newPremises}, and a map \textit{varmap}. Here, \textit{newPremises} is the same set of premises $P$ in which each variable has been given fresh, distinct names, if occurring multiple times. \textit{varmap} maps each of the variables that have been replaced to the set of new variables that replaced them. The reason for collecting these new variables in \textit{varmap} is because we need to relate them by subtyping. (This is the responsibility of the second step, which works based on the information in \textit{varmap}.)

To make an example, when $\proc{Split-Equal-Types}$ is applied to the premises of \textsc{(t-app)}, we have 

\noindent Input: $P = \{\Gamma \typeOf \app e_1 : T_{1}\to T_2$, $\Gamma \typeOf \app e_2 : T_{1}\}$\\
Output = $(\textit{newPremises}, \textit{varmap})$ where 
\[
\begin{array}{rl}
\textit{newPremises}&=\{\Gamma \typeOf \app e_1 : T_{11}\to T_2, \Gamma \typeOf \app e_2 : T_{12}\} \\
\textit{varmap}&=\{T_1 \mapsto \{T_{11}, T_{12}\}\}
\end{array}
\]

$\proc{Split-Equal-Types}$ produces this output in the following way. 
Line 1 initializes \textit{newPremises} to the empty set, and \textit{varmap} to the empty map. 
The loop at lines 2-8 is executed for each premise $p$ of the set of premises $P$. 
For example, with \textsc{(t-app)} we have two iterations; the first is with $p = \Gamma \typeOf \app e_1 : T_{1}\to T_2$, and the second is with $p = \Gamma \typeOf \app e_2 : T_{1}$. Line 3 extracts the components of the typing premise. It does so in a style that is reminiscent of pattern-matching. 
The component that is relevant for the algorithm is $\textit{someType}$, which is the output type of the typing premise. 
The loop at lines 4-7 applies to each variable $T$ in $\textit{someType}$ that appears more than once in the premises of $P$. 
We focus on variables that have multiple occurrences because variables that occur only once do not need to be replaced with new names. 
For each of these variables $T$, we generate a fresh variable that is not used in $P$. We do so with $\proc{Fresh}(P)$ at line 5. Line 6 modifies the premise $p$ by rewriting it to use the fresh variable in lieu of $T$. Line 7 also updates \textit{varmap} to add the fresh variable to the set of new variables mapped by $T$. 
Line 8 adds $p$ to \textit{newPremises}. At that point, $p$ may have been modified with line 6, or may have remained unchanged.  
Finally, line 9 returns the pair $(\textit{newPremises},\textit{varmap})$.

\paragraph{Relate New Variables by Subtyping} 
This second step is performed in the context of the procedure $\proc{Add-Subtyping}$. This is our general procedure that takes a language specification in input, and adds subtyping to it. To do so, $\proc{Add-Subtyping}$ calls \proc{Split-Equal-Types}, and then works on the rules modified by \proc{Split-Equal-Types} to relate the new variables by subtyping. 

The pseudo-code of $\proc{Add-Subtyping}$ is the following. 

{
\begin{codebox}
  \Procname{$\proc{Add-Subtyping}(L)$}
  \li \For $\text{each rule } r \in L.rules \text{ s.t. }r.conclusion \text{ is of the form } \Gamma \vdash e : \textit{someType}$ \Do
  \li $(\textit{newPremises},\textit{varmap}) \gets \proc{Split-Equal-Types}(r.premises)$
  \li \For $\text{each mapping } (T \mapsto \textit{setOfNewVars}) \text{ in } \textit{varmap}$ \Do
  \li \If $\text{there exists a type in \textit{setOfNewVars} that is invariant in \textit{newPremises}}$ \Do
  \li \For $\text{each } T_1,T_2 \in \textit{setOfNewVars}$ \Do
  \li $\textit{newPremises} \gets \textit{newPremises} \cup \{T_1 = T_2\}$
  \End
  \li \ElseIf $\text{there is exactly one type $T'$ in \textit{setOfNewVars} that is contravariant in \textit{newPremises}}$ \Do
  \li \For $\text{each } T_{new} \in (\textit{setOfNewVars} \setminus {T'})$ \Do
  \li $\textit{newPremises} \gets \textit{newPremises} \cup \{T_{new} <: T'\}$
  \End
  \li \ElseIf $\text{none in \textit{setOfNewVars} is contravariant or invariant in \textit{newPremises}}$ \Do
  \li $\text{say that } \textit{setOfNewVars}  = \{T_1, T_2,  \ldots, T_n\}$
  \li $\textit{newPremises} \gets \textit{newPremises}  \cup \{T = T_1 ~\vee ~T_2 ~\vee ~\ldots ~\vee ~T_n\}$
  \li $\textit{newPremises} \gets \textit{newPremises}  \cup \{T_{new} <: T'\}$
\Else $error$
  \End
  \End
  \li $r.premises \gets \textit{newPremises}$
\end{codebox}
}
The argument $L$ is the language specification in input. 
The procedure modifies the rules of $L$ in-place. 
Line 1 selects only the typing rules of $L$ (leaving out reduction rules, for example). 
It does so by selecting only the rules whose conclusion has the form of a typing formula. 
Lines 2-14 constitute the body of the loop, and apply for each of these rules.  
Line 2 calls $\proc{Split-Equal-Types}$, passing the premises of the typing rule as argument. This call returns the new premises and the map previously described. 
Lines 3-13 iterate over the key-value pairs of the map. Key-value pairs are of the form $T \mapsto \textit{setOfNewVars}$, where $T$ is the variable that occurred in the original typing rule before calling $\proc{Split-Equal-Types}$. We dub $T$ as the \emph{original variable}. Also, $\textit{setOfNewVars}$ contains the new variables generated by $\proc{Split-Equal-Types}$ for $T$. 

Lines 4-6 cover the case for when the original variable appeared in invariant position. 
In that case, there exists a variable in \textit{setOfNewVars} that is in invariant position in \textit{newPremises}, which we check with line 4. 
As the original variable appeared in invariant position, all the new variables must be related by equality. (We make an example shortly). Therefore, lines 5-6 add an equality premise for every two variables in \textit{setOfNewVars}. 
This case covers operators such as the assignment in a language with references, as $T$ is invariant in a reference type $\key{Ref} ~ T$. 
Consider the typing rule for the assignment operator on the left, and its version with subtyping on the right. 

{
\begin{gather*}
{
\ninference{t-assign}
	{
	\Gamma \typeOf \app e_1 : \key{Ref} ~ T \\
	\Gamma \typeOf \app e_2 : T	
	} 
	{ \Gamma \typeOf \app e_1 := e_2 : \key{unitType}}
}
\quad \Longrightarrow \quad 
{
\ninference{t-assign'}
	{
	\Gamma \typeOf \app e_1 : \key{Ref} ~ T_1 \\
	\Gamma \typeOf \app e_2 : T_2 \\ T_1 = T_2
	} 
	{ \Gamma \typeOf \app e_1 := e_2 : \key{unitType}}
}
\end{gather*}
}

Here, $\proc{Split-Equal-Types}$ replaces $T$ with two new variables $T_1$ and $T_2$, but as $T$ is invariant in \textsc{(t-assign)}, we generate the premise $T_1 = T_2$, which is the correct outcome. 

Lines 7-9 cover the case for when the original variable appeared in a contravariant position. 
In that case, there exists a type $T'$ in \textit{setOfNewVars} that is contravariant in \textit{newPremises}. We detect such a case with line 7. 
Notice that line 7 also checks that the original variable appeared only once in contravariant position. We address this aspect later when we discuss line 13. 
As $T'$ appears in contravariant position, this is an input that is waiting to receive values. 
Therefore, we generate the subtyping premises that set all the other new variables in \textit{setOfNewVars} as subtypes of $T'$ (lines 8-9).  
This case covers operators such as the function application, which we have discussed previously. 
Thanks to lines 7-9, $\proc{Add-Subtyping}$ generates the typing rule \textsc{(t-app')} from \textsc{(t-app)}, which is the correct outcome. 

Lines 10-12 cover the case in which variance does not play a role. In this case, all the newly generated variables are peers. (We will make an example shortly). 
Therefore, we compute the join $\vee$ for them \cite{tapl}. 
This case applies to operators such as if-then-else. Consider the typing rule of if-then-else below on the left, and its version with subtyping on the right. 

{
\begin{gather*}
{
\ninference{t-if}
  {
  \Gamma \typeOf e_1 : \Bool \qquad \Gamma \typeOf e_2 : T \\ \Gamma \typeOf e_3 : T
  }
  {\Gamma \typeOf (\textit{if} \app e_1 \app e_2 \app e_3) : T}
}
\quad \Longrightarrow \quad 
{
\ninference{t-if'}
  {
  \Gamma \typeOf e_1 : \Bool \qquad \Gamma \typeOf e_2 : T_1 \\ \Gamma \typeOf e_3 : T_2 \\\\ T = T_1 \vee T_2
  }
  {\Gamma \typeOf (\textit{if} \app e_1 \app e_2 \app e_3) : T}
}
\end{gather*}
}

Here, $\proc{Split-Equal-Types}$ replaces $T$ with two new variables $T_1$ and $T_2$. Then, line 10 detects that variance does not play a role for these new variables. 
Indeed, the two branches of the if-then-else are peers. Therefore, lines 11 and 12 generate the premise that computes the join of all the new variables, and assign it to $T$. 
Thanks to lines 10-12, \textsc{(t-if')} is precisely the typing rule that $\proc{Add-Subtyping}$ generates, which is the correct outcome. 
Another example where variables are peers is with the \key{case} operator of the sum type. 

Line 13 throws an error if none of the previous cases apply. This happens, for example, if a variable appears in contravariant position multiple times in the typing rule. 
Consider the following typing rule.

{
\begin{gather*}
\inference{
  \Gamma \vdash e_1 : T \to (T \to \Bool) \qquad \Gamma \vdash e_2 : T \times T
}{\Gamma \vdash \key{ app2}\; e_1\; e_2 : \Bool}
\end{gather*}
}

Here, $T$ appears in contravariant position twice in the type of $e_1$. However, the typing rule of \key{ app2} cannot distinguish how the components of the pair $e_2$ are going to be used. Consider two alternative reduction rules for \key{ app2}: 
{
  \begin{gather*}
    \key{ app2}\; e_1\; e_2\step  ((e_1\app (\key{fst}\; e_2)) \app (\key{snd}\; e_2))  \qquad or \qquad \key{ app2}\; e_1\; e_2\step  ((e_1\app (\key{snd}\; e_2)) \app (\key{fst}\; e_2))
  \end{gather*}
}

The reduction rule on the left entails that the first component of the pair $e_2$ must be subtype of the first $T$ of $T \to (T \to \Bool)$, and that the second component of the pair $e_2$ must be subtype of the second $T$ of $T \to (T \to \Bool)$. 
Conversely, the reduction rule on the right entails that the second component of the pair $e_2$ must be subtype of the first $T$ of $T \to (T \to \Bool)$, and that the first component of the pair $e_2$ must be subtype of the second $T$ of $T \to (T \to \Bool)$. 

However, $\proc{Add-Subtyping}$ only analyzes the typing rule of \key{ app2}, which alone is not informative enough to tell about the parameter passing to $e_1$. 
Therefore, we do not know what subtyping premises to generate. In this case, $\proc{Add-Subtyping}$ throws an error. 

To solve this problem, we could extend $\proc{Add-Subtyping}$ to analyze the reduction semantics of \key{ app2}, but we observe that language designers may specify such semantics in a way that is as complex as they wish. Reduction rules may not use parameter passing immediately and evidently, in favor of jumping from operator to operator several times, which makes the analysis hard to do. 
For these reasons, we have not investigated this path, also because we may be speaking about cases that are quite uncommon, and not strictly necessary to cover in detail in our undergraduate class.

Finally, line 14 replaces the premises of $r$ with $\textit{newPremises}$. 
The relations $\vee$ and $<:$ can be generated with an algorithm, too, but we omit showing these procedures here. 
In this paper, we simply want to illustrate the approach rather than strive for completeness. 
%
\subsection{Language Transformation for CK}\label{ck}

We have taught abstract machines following the notes of Felleisen and Flatt \cite{FelleisenFlatt02}. 
To recap, the CK machine remedies an inefficiency aspect of the reduction semantics. Consider the following reductions:\\
(\key{hd} retrieves the head of a list, \key{t} and \key{f} are the constants for the true and false boolean, respectively). 
{
\begin{align*}
  (\key{if}\; (\key{hd}\; [\key{f}\land \HI{$((\lambda x.x) \app \key{t})$},\key{t}]))\; e_1\; e_2) &\step (\key{if}\; (\key{hd}\; [\key{f}\land  \key{t},\key{t}]))\; e_1\; e_2)\\
                                                                                (\key{if}\; (\key{hd}\; [\HI{$\key{f}\land  \key{t}$},\key{t}]))\; e_1\; e_2) &\step \ldots 
\end{align*}
}
To perform the step at the top, the reduction semantics traverses the term and seeks for the first available evaluation context, which points to the highlighted subterm. 
At the second step, the reduction semantics must seek again for an available evaluation context, and does so by traversing the term again from the top level \key{if} operator, which is inefficient. 

To improve on this aspect, and avoid these recomputations, the CK machine carries a \textit{continuation} data structure at run-time. 
The grammar for continuations, and the CK reduction rules for function application are the following. \\
(\key{mt} is the empty continuation, which denotes machine termination.)
\begin{syntax}
  \text{Continuation} & k & ::= & \key{mt}  \mid (\key{app}_1\app  e \app k) \mid (\key{app}_2\app  v \app k)
\end{syntax}
  \vspace{-2em}
\begin{align*}
(\key{app} \app e \app e_2), k & \step e, (\key{app}_1\app  e_2 \app k)& \key{Start}\\
v, (\key{app}_1 \app e \app k) & \step  e, (\key{app}_2 \app v \app k) & \key{Order} \\
v,  (\key{app}_2 \app (\lambda x.e)\app k) & \step e[v/x], k& \key{Computation}
\end{align*}

The reduction relation has the form $e,k \step e,k$, where $k$ is built with continuation operators \key{mt}, $\key{app}_1$, and $\key{app}_2$. 
There is a continuation operator for each evaluation context. 
Each continuation operator has always an argument $k$, which is the next continuation, and one expression argument less than the operator because one of the expressions is currently out to be the focus of the evaluation. 
For example, $(\key{app}_2 \app v \app k)$ means that the current expression being evaluated returns as the second argument of the application, and $v$ is the function waiting for such argument. 

Below, we show the language transformations for generating the CK machine, except for the procedure that generates the \key{Computation} rule above, because that procedure is straightforward. 

\paragraph{Generating the Grammar for Continuations}

The following pseudo-code generates the grammar \key{Continuation}. 

{
\begin{codebox}
  \Procname{$\proc{ck-generate-grammar}(\textit{EvalCtx})$}
  \li $\text{create grammar category } \key{Continuation} \text{, and add grammar item } \key{mt} \text{ to it}$ 
  \li \For $\text{each } (\key{op}\; t_1 \ldots t_n) \in \textit{EvalCtx}$ \Do
  \li \If $t_i = E$ \Do
  \li $\text{add grammar item } (\key{op}_i\; (t_1 \ldots t_n \text{ minus } E) ~ k) \text{ to } \key{Continuation}$  
  \End
\end{codebox}
}

For each evaluation context, the index where the $E$ appears determines the index of the continuation operator. 
The arguments of this operator are all the arguments that are not $E$. Indeed, the argument at that position will currently be the focus of the evaluation. 
Also, the next continuation $k$ is the last argument. 

\paragraph{Generating the \key{Start} rule}

The following pseudo-code generates the reduction rule \key{Start}, which brings the computation of an operator into using continuation operators. 

{
\begin{codebox}
  \Procname{$\proc{ck-generate-start}(\textit{Continuations})$}
  \li $\text{find }  (\key{op}_i\; t_1 \ldots t_n ~ k) \in \textit{Continuations} \text{ with no $v$}$ 
  \li $\text{add reduction rule }  (\key{op}\; t_1 \ldots ~ e ~\ldots t_n), k \step e, (\key{op}_i\; t_1 \ldots t_n ~ k)$
  \End
\end{codebox}
}

Here, $e$ appears at position $i$ in \key{op}. 
If a continuation contains some $v$ as arguments, it means that those arguments must have been the subject of some other evaluation context that evaluated them to a value. 
Therefore, that cannot be the starting point. Our starting point, instead, is a continuation that contains no $v$. 
The reduction rule that we add takes the operator into using the continuation operator that we have just found. 

\paragraph{Generating \key{Order} rules}
The following pseudo-code generates the reduction rules \key{Order}. These rules evaluate the arguments of the operator by jumping from one continuation operator to another in the order established by the evaluation contexts. 

{
\begin{codebox}
  \Procname{$\proc{ck-generate-order}(\textit{Continuations},\textit{EvalCtx})$}
  \li \For $\text{each }  (\key{op}_i\; t_1 \ldots t_m ~ k) \in \textit{Continuations}$ \Do
  \li $\text{find }  (\key{op}\; t_1' \ldots t_n') \in \textit{EvalCtx} \text{ where } (t_k = t_k'  \text{ or } t_k' = E, \text{ for all $k$ })$ 
  \li \If $t_j' = E$ \Do
  \li $\text{add reduction rule }  v, (\key{op}_i\; t_1 \ldots t_m ~ k) \step t_j, (\key{op}_j\; t_1 \ldots ~v~ \ldots t_m ~ k) $
  \End
\end{codebox}
}

Here, $v$ appears at position $i$ in $\key{op}_j$. 
After finishing an evaluation in the contexts of the continuation $\key{op}_i$, we need to find the next continuation operator $\key{op}_j$. To do so, we find a match between the arguments of the continuation $\key{op}_i$ with arguments of an evaluation context. 
This is because arguments that are values in the continuation then need to be values, too, in the next evaluation context. Arguments that are simply expressions $e$ in the continuation then need to be expressions $e$, too, in the next evaluation context. 
The evaluation context will have, however, an argument $E$ (and only one argument $E$) at some position $j$, which identifies the index of the next continuation operator. 
The reduction rule that we add starts from a point where a value has been computed, and we are in the context of the continuation $\key{op}_i$. In one step, we extract the $j$-th argument of the continuation $\key{op}_i$ because that is the expression that now needs to be in the focus of the evaluator. The next continuation is then $\key{op}_j$, where we placed the value $v$ just computed among the arguments of $\key{op}_j$, and specifically at position $i$.

Generating \key{Computation} rules is rather straightforward, hence we omit showing that simple procedure.   

\subsection{Final Exam}\label{exam}

At the end of the course, students have been evaluated with a final exam. The final exam included questions about subtyping and CK machines\footnote{The final exam also contained questions about other topics of the course. For example, the final exam of the second iteration contained questions about garbage collection. However, here we focus only on the parts of the exam that concern language transformations.}. 
The goal is not to test students on the language transformations per se, but rather on their understanding of subtyping and the CK machine. 
We therefore tested whether students would be able to use their understanding in practice. 
Our questions tested students on whether, if presented with a language with unusual operators, they would be able to add subtyping to it, and derive its CK machine. 

We have delivered two iterations of the course. The final exam took place online on both iterations due to the COVID-19 pandemic. 
In the first iteration of the course, we have shared a link to a text file that contained the content of the exam, and students submitted an updated text file via email. In the second iteration, the text of the exam was uploaded in the Blackboard system\footnote{\url{https://www.blackboard.com/}}. Students could insert their answers on the webpage as text, and submit them with the submit button. 

The text of the final exam had two parts: 

\begin{itemize}
\item The description of a toy language called $\toylanguage$. 
\item The questions that students were asked to answer, which referred to the language $\toylanguage$.
\end{itemize}

Below, we describe these two parts.

\paragraph{The Toy Language $\toylanguage$}

The text of the exam contained a description of $\toylanguage$. 
The text told the students that $\toylanguage$ is a $\lambda$-calculus with pairs $\langle e_1, e_2 \rangle$ and lists $[e_1, \ldots, e_n]$, equipped with two operators called $\doublyApply$ and $\addToPairAsList$, which we describe below. The text of the exam did not repeat the typing rules and reduction rules of the $\lambda$-calculus with pairs and lists because we have seen them extensively in class, and because they did not play a role in the questions of the exam. On the contrary, the text of the exam provided the students with the formal semantics of $\doublyApply$ and $\addToPairAsList$, which we will show shortly. 

Below, we describe the operators $\doublyApply$ and $\addToPairAsList$.

\begin{itemize}
\item $\doublyApply$: The text of the final exam contained the following description of $\doublyApply$. 
``$\doublyApply$ takes two functions $f_1$ and $f_2$ in input, and two arguments $a_1$ and $a_2$, and creates the pair $\langle f_2(f_1(a_1)), f_1(f_2(a_2))\rangle$. 
That is, the first component of the pair calls $f_1$ with $a_1$ and passes the result to $f_2$, and the second component calls $f_2$ with $a_2$ and passes the result to $f_1$.''

The text of the exam also provided the students with the following syntax, evaluation contexts, typing rule, and reduction rule for $\doublyApply$. 
\begin{syntax}
  \text{Expression} & e & ::= & \ldots \mid (\doublyApply \app e \app e \app e \app e) \\
  \text{Evaluation Context} & E & ::= & \ldots  \mid (\doublyApply \app E \app e \app e \app e)   \mid (\doublyApply \app v \app E \app e \app e) \\
  &&&  \mid (\doublyApply \app v \app v \app E \app e)   \mid (\doublyApply \app v \app v \app v \app E)   
\end{syntax}
\begin{gather*}
\inference{
\Gamma \vdash e_1 : T_1 -> T_2 &
\Gamma \vdash e_2 : T_2 -> T_1\\
\Gamma \vdash e_3 : T_1&
\Gamma \vdash e_4 : T_2
}
{
\Gamma \vdash (\doublyApply \app e_1 \app e_2 \app e_3 \app e_4) : T_2 \times T_1
}
\\[2ex]
 \doublyApply \app v_1 \app v_2 \app v_3 \app v_4 \step  \langle(v_2 \app (v_1\app v_3)), (v_1 \app (v_2\app v_4))\rangle
\end{gather*}
%
\item $\addToPairAsList$: The text of the exam contained the following description of $\addToPairAsList$.  
``$\addToPairAsList$ takes an element $a_1$ and a pair $p$, and strives to add the element to the pair. As pairs contain only two elements, it creates a list with three elements: the element $a_1$, the first component of $p$, and the second component of $p$.''  To make a concrete example, we have $\addToPairAsList \,\app a_1 \app \langle a_2,a_3\rangle = [a_1, a_2, a_3]$.  

The text of the exam also provided the students with the following syntax, evaluation contexts, typing rule, and reduction rule for $\addToPairAsList$. 
\begin{syntax}
  \text{Expression} & e & ::= & \ldots \mid  (\addToPairAsList \app e \app e)  \\
  \text{Evaluation Context} & E & ::= & \ldots  \mid (\addToPairAsList \app E \app e)   \mid (\addToPairAsList \app v \app E ) 
\end{syntax}
\begin{gather*}
\inference{
\Gamma \vdash  e_1 : T &
\Gamma \vdash  e_2 : T \times T
}
{
\Gamma \vdash  (\addToPairAsList \app e_1\app e_2) : List \app T
}
\\[2ex]
\addToPairAsList \app v_1 \app\langle v_2,v_3\rangle \step [v_1, v_2, v_3]
\end{gather*}
\end{itemize}
Although these operators are not extremely bizarre, it is unusual to see them as primitive operations. 

\paragraph{Questions and their Challenges}

After the description of the language $\toylanguage$, the text of the exam gave the students three questions that they were asked to answer. 
We dub these questions ``Subtyping of $\doublyApply$'', ``Subtyping of $\addToPairAsList$'', and ``CK for $\doublyApply$'', respectively. 

The question ``Subtyping of $\doublyApply$'' asked the students to show the version of the typing rule of $\doublyApply$ with subtyping. 
This task is not trivial because the typing rule of $\doublyApply$ has three occurrences of $T_1$, and one of them is in contravariant position, which is the input of a function. 
Therefore, the other two occurrences of $T_1$ must be subtypes of that. The same scenario occurs for $T_2$. 

The correct answer to this question is the following: \\
(The output type of this typing rule is more restrictive than necessary. The output type could be adjusted by applying another procedure, but we have omitted this part). 
\begin{gather*}
\inference{
\Gamma \vdash e_1 : T_1 -> T_2' &
\Gamma \vdash e_2 : T_2 -> T_1'\\
\Gamma \vdash e_3 : T_1'' &
\Gamma \vdash e_4 : T_2'' \\ 
T_1' <: T_1 & T_1'' <: T_1 &
T_2' <: T_2 & T_2'' <: T_2 
}
{
\Gamma \vdash (\doublyApply \app e_1 \app e_2 \app e_3 \app e_4) : T_2 \times T_1
}
\end{gather*}

The question ``Subtyping of $\addToPairAsList$'' asked the students to show the typing rule of the operator $\addToPairAsList$ with subtyping added. 
This task is also non-trivial because there are three occurrences of $T$ that are peers. Therefore, the correct solution is to compute a join type. 

The correct answer to this question is the following: 

\begin{gather*}
\inference{
\Gamma \vdash  e_1 : T' &
\Gamma \vdash  e_2 : T'' \times T'''\\
T = T' \vee T'' \vee T'''
}
{
\Gamma \vdash  (\addToPairAsList \app e_1\app e_2) : List \app T
}
\end{gather*}

The question ``CK for $\doublyApply$'' asked the students to derive the CK machine for $\toylanguage$ insofar as the reduction rules for $\doublyApply$ are concerned. 
This operator is challenging because it has a high number of arguments (four). To complete the task, students must understand well the relationship between continuations and the evaluation order of arguments. 

The correct answer to this question is the following: 

{
\begin{align*}
(\doublyApply\; e_1\; e_2\; e_3\; e_4), k \step e_1, (\doublyApply_1\; e_2\; e_3\; e_4\; k) & \qquad \key{Start}\\
v_1, (\doublyApply_1\; e_2\; e_3\; e_4\; k) \step e_2, (\doublyApply_2\; v_1\; e_3\; e_4\; k) & \qquad \key{Order}\\
v_2, (\doublyApply_2\; v_1\; e_3\; e_4\; k) \step e_3, (\doublyApply_3\; v_1\; v_2\; e_4\; k)& \qquad \key{Order}\\
v_3, (\doublyApply_3\; v_1\; v_2\; e_4\; k) \step e_4, (\doublyApply_4\; v_1\; v_2\; v_3\; k)& \qquad \key{Order}\\
v_4, (\doublyApply_4\; v_1\; v_2\; v_3\; k)\step \langle(v_2\; (v_1\; v_3)), (v_1\; (v_2\; v_4))\rangle, k & \qquad \key{Computation}
\end{align*}
}

The exam could also ask for the CK reduction rules of $\addToPairAsList$. However, this task is slightly simpler than $\doublyApply$, and we therefore were not interested in requesting those rules. 

\section{Evaluation}


As we have previously said, we have run two iterations of the undergraduate PL course that we have described. 
To evaluate the merits of our thesis, we have collected information about students' success with the final exam, and more specifically, with their success in answering the questions ``Subtyping of $\doublyApply$'', ``Subtyping of $\addToPairAsList$'', and ``CK for $\doublyApply$''. 


For each question, we have evaluated the answer of each student as ``Correct'', ``Partially Correct'', ``Partially Incorrect'', and ``Incorrect/Missing''. 
Students' answers were classified as ``Correct'' only if they matched the solution given in the previous section. 
Answers were classified as ``Incorrect/Missing'' if they were missing, or they were completely incorrect. What constitutes a completely incorrect, a partially incorrect, and a partially correct answer is subjective by nature, therefore we have to draw a line in the sand, somehow subjectively. Our rationale is the following. A ``Partially Correct'' answer does not match the solution but shows that the student was on the way towards a correct solution. 
A ``Partially Incorrect'' answer contains some elements that demonstrates that the student is applying some correct reasoning principles. 
A completely incorrect answer (``Incorrect/Missing'') provides no indication that the student is applying correct reasoning principles. 


In total, we have conducted the study on 55 students. 
The rating of students' answers is shown in the following table. 
 \begin{center}
\begin{table}[hbt!]
{
\qquad
\begin{tabular}{lllll}
\hline
\multicolumn{1}{|l|}{} & \multicolumn{1}{l|}{Correct} & \multicolumn{1}{l|}{$\begin{array}{c}\text{Partially} \\ \text{Correct}\end{array}$} & \multicolumn{1}{l|}{$\begin{array}{c}\text{Partially} \\ \text{Incorrect}\end{array}$}  & \multicolumn{1}{l|}{Incorrect/Missing} \\ \hline
\multicolumn{1}{|l|}{Subtyping of $\doublyApply$} & \multicolumn{1}{l|}{15} & \multicolumn{1}{l|}{11} & \multicolumn{1}{l|}{15} & \multicolumn{1}{l|}{14} \\ \hline
\multicolumn{1}{|l|}{Subtyping of $\addToPairAsList$} & \multicolumn{1}{l|}{22} & \multicolumn{1}{l|}{10} & \multicolumn{1}{l|}{11} & \multicolumn{1}{l|}{12} \\ \hline
\multicolumn{1}{|l|}{CK for $\doublyApply$} & \multicolumn{1}{l|}{18} & \multicolumn{1}{l|}{13} & \multicolumn{1}{l|}{10} & \multicolumn{1}{l|}{14} \\ \hline
\end{tabular}
}
\end{table}
\end{center}

The question ``Subtyping of $\doublyApply$'' seems to be the most difficult among the three, as shown by the lowest number of completely correct answers. Subtyping of $\doublyApply$ is indeed a rather complicated task, as it involves contravariance. Furthermore, many variables are around, and a good number of them need to be subtype of a same variable. It is not surprising that 29 out of 55 did not provide a good answer (and were ``Partially Incorrect'' or ``Incorrect/Missing''). On the contrary, it is rather encouraging to see that 26 out of 55 students could provide a good answer (``Correct'' or ``Partially Correct''). 

The question ``Subtyping of $\addToPairAsList$'' seems to be the easiest among the three, as the highest number of students could provide a completely correct answer. Students could detect more easily that types are treated as peers in the typing rule of $\addToPairAsList$, and perhaps this signals that this case is simpler to grasp than the contravariant case of $\doublyApply$. 

We are surprised by the results of the question ``CK for $\doublyApply$'', as such a machine seems to be rather involved. Regardless, a good number of students (18) could provide a completely correct answer, and a high number of them could give a good answer (``Correct'' or ``Partially Correct''). This is indicative that students could grasp the mechanics of the evaluation order, and translate it well as a CK machine. 

It is safe to imagine that most students have not been exposed to formal semantics until this very course, and these questions are generally hard for them. It is encouraging to see that a good number of students could provide good answers. It may be an indication that, by and large, students could gain an understanding of subtyping and CK machines. However, we would like to explicitly say that we do not draw any general conclusion from this data. 

\paragraph{A Note on Correctness} 
As we have previously said, we do not claim any theoretical results of correctness of the algorithms that we have taught in class. However, we have implemented them as tools (\cite{lnc}) that take a language specification in input written as a textual representation of operational semantics (with syntax similar to that of Ott \cite{Sewell:2007}), and output the modified language specification (in the same textual format). 
We have applied these tools to several functional languages in order to add subtyping to them and derive their CK machines, and we have confirmed by inspecting the output languages that we have obtained the correct formulations.

\subsection{Threats to Validity}

The following observations keep this paper from drawing general conclusions about the thesis that language transformations are beneficial in class.  

\paragraph{Further Studies}

While 55 students is a decent number, we would like to conduct more iterations of the same course, and have a larger pool of participants. 
When more data will be gathered, we plan to report on such data in a journal version of this paper. 

\paragraph{Negative Experiments?}

It would be interesting to run instances of the course with language transformations, and also run instances \emph{without} language transformations, while keeping the same syllabus and the final exam. The goal is to see whether there is a significant improvement in the success rate of exams in those courses that have used language transformations. 
However, we find pedagogical issues in implementing this plan. 
We think that adopting the final exam of Section 2.3 without having taught language transformations may not be a sensible choice. 
For example, simply covering subtyping with TAPL may not provide students with sufficient knowledge to complete the exam, and we may put unrealistic expectations on students' ability to generalize and extrapolate general programming languages principles at the undergraduate level. 

\paragraph{Perceived Effectiveness?}
We have made an attempt to evaluate whether students perceived that using language transformations was helpful for their learning. 
At the end of the course, we have given a survey for them to fill in. 
The survey contained six statements which, as typical in surveys, required a rating. 
For example, to evaluate the task for ``Subtyping of $\doublyApply$'', the survey had the statements: ``The language transformation algorithm for adding subtyping to languages helped me understand subtyping better'', and ``The language transformation algorithm for adding subtyping to languages helped me add subtyping to the language at hand during the exam''. Students could assign a grade among ``Strongly Agree'', ``Somewhat Agree'', ``Neither Agree nor Disagree'', ``Somewhat Disagree'' or ``Strongly Disagree'' to the statement. The survey requested students to rate the equivalent statements for ``Subtyping of $\addToPairAsList$'' and ``CK for $\doublyApply$''. 

Unfortunately, the survey did not receive participation. Our courses have taken place virtually during the COVID-19 pandemic, which may have been the cause of the experienced lack in participation. 

\section{Future Work}

In this section, we discuss our plans. Our first goal is to evaluate the perceived effectiveness of language transformations with the survey that we have just described. 
Hopefully, participation to the survey will improve in the future. 
Other venues for future work are the following. 

\paragraph{Improving our Current Language Transformations}

The procedures of Section 2.2 produce CK machines without environments, which is not typical. Our next step is to extend our procedures to capture the full Felleisen and Friedman's
CEK machine. 
Similarly, we plan on developing language transformations to automatically derive other popular abstract machines such as Landin's SECD \cite{Landin65}, and Krivine's KAM \cite{krivine2007call} machines.

The language transformation that we have used for subtyping works only on simple types (sums, products, options, etcetera). 
We would like to extend the algorithm to capture also constructors that carry maps, such as records and objects. Maps can associate field names to values, and method names to functions. 
Maps come with their own subtyping properties such as width-subtyping, and permutation \cite{tapl}, and we plan on extending our algorithm to cover them. 
Similarly, we would like to develop language transformations for automatically adding bounded polymorphism \cite{CaMiMaSc1994,Abadi:1996}, recursive subtyping \cite{Amadio:1993}, multiple inheritance and mixins \cite{Bracha92,Cardelli84} to languages, to make a few examples. 
\paragraph{Language Transformations for Other Features}
Subtyping and abstract machines are not the only features that can be taught in a course in the principles of programming languages. 
We plan on addressing other features with language transformations, and using them in class. 

In teaching the formalism of operational semantics, instructors may begin with a small-step or with a big-step semantics style. 
Whichever style has been chosen, it could be beneficial to explain the other style with language transformations (in addition to the planned material)
that turn small-step into big-step, or big-step into small-step, respectively. 
Much work has been done to translate one style into the other \cite{Ciobaca13,danvy:reduction-free,Danvy:2008,danvy2004refocusing}\footnote{Among these works, \cite{Ciobaca13} seems to be the most suitable for language transformations.}, and we plan on building upon this work. 
It is worth noting that converting from one style to the other may come with limitations. For example, it may not be possible to derive an equivalent big-step semantics from a small-step semantics formulation when parallelism is involved. The mentioned translation methods are subject to these limitations, and so will our corresponding language transformations.  

We would like to develop a language transformation for automatically generating Milner-style type inference procedures.  
Also, we would like to devise a language transformation for adding generic types to languages. 
Some courses teach dynamic typing and run-time checking in some detail. We would like to explore the idea of automatically generating the dynamic semantics of dynamically typed languages based on a given type system. 
That is, a language transformation which relies on the type system to inform how the dynamic semantics should be modified in order to perform run-time type checking. 

We are not aware of any work that automates the adding of the latter three examples. 
Developing such language transformations may be challenging research questions on their own. 

\paragraph{Advanced Tasks}
Language transformations may be integrated in graduate level courses, as well. 
Some of these courses have a research-oriented flavour. 
In such courses, instructors may assign advanced tasks with language transformations. 
For example, instructors may ask students to study the work of Danvy et al. \cite{danvy:reduction-free,Danvy:2008,danvy2004refocusing} to derive reduction semantics. 
The approach is rather elaborate, and involves techniques such as refocusing and transition compression. 
Instructors may ask students to develop a series of language transformations that capture this method. 
Similarly, instructors may ask students to model the language transformation for generating the pretty-big-step semantics from a small-step semantics \cite{PoulsenM14}.  
Another idea is to target the Gradualizer papers \cite{Cimini:2016aa,Cimini:2017}, for automatically adding gradual typing to languages. 

\section{Conclusion}
Instructors can integrate language transformations 
into their undergraduate PL courses. 
We do not advocate replacing material, but to use language transformations in addition to the planned material. 
Our thesis is that language transformations are beneficial for students to help them deepen their understanding of the PL features being taught. 
In this paper, we have presented the study that we have conducted, and the results from this study. 
Although we refrain from declaring language transformations unequivocally beneficial, our numbers are encouraging, and we also offer this paper to open a conversation on the topic, and to inspire similar studies towards gathering evidence for, or against, our thesis.  

\paragraph{Acknowledgements}
We would like to thank our EXPRESS/SOS 2021 reviewers for their feedback, which helped improve this paper.

\bibliographystyle{eptcs}
\bibliography{biblio.bib} 

\end{document}

%% file: listing-macros.tex
\colorlet{lprolog}{blue!70!black}
\colorlet{abellatop}{blue!70!green}
\colorlet{abellatac}{orange!30!black}
\colorlet{abellabad}{red!80!yellow}

\lstdefinelanguage{lprolog}{%
  alsoletter={-},
  classoffset=0,%
  morekeywords={sig,module,type,kind,pi,sigma,end},%
  keywordstyle=\color{lprolog},%
  classoffset=0,%
  otherkeywords={:-,=>,<=,\&},%
  sensitive=true,%
  morestring=[bd]",%
  morecomment=[l]\%,%
  morecomment=[n]{/*}{*/},%
}

\lstdefinelanguage{abella}[]{lprolog}{%
  alsoletter={-},
  classoffset=1,%
  morekeywords={Close,CoDefine,Define,Kind,Query,Quit,Specification,
    Set,Split,Theorem,Type,Undo,by,as,prop,true,false,forall,exists,nabla},%
  keywordstyle=\color{abellatop},%
  classoffset=2,%
  morekeywords={abbrev,apply,backchain,case,coinduction,cut,
    induction,inst,intros,monotone,on,permute,rename,left,right,witness,
    search,split,to,unabbrev,unfold,assert,with},%
  keywordstyle=\color{abellatac},%
  classoffset=3,%
  morekeywords={undo,abort,skip,clear},%
  keywordstyle=\color{abellabad}\underbar,%
  classoffset=0,%
}

%% file: definitions.tex
\newcommand*\colourcheck[1]{%
  \expandafter\newcommand\csname #1check\endcsname{\textcolor{#1}{\text{\ding{52}}}}%
}
\colourcheck{blue}
\colourcheck{green}
\colourcheck{red}

\newcolumntype{Y}{>{\raggedleft\arraybackslash}X}

\tcbset{tab1/.style={fonttitle=\bfseries\large,fontupper=\normalsize\sffamily,
colback=yellow!10!white,colframe=red!75!black,colbacktitle=red!40!white,
coltitle=black,center title,freelance,frame code={
\foreach \n in {north east,north west,south east,south west}
{\path [fill=red!75!black] (interior.\n) circle (3mm); };},}}

\tcbset{tab2/.style={enhanced,fonttitle=\bfseries,fontupper=\small\sffamily,
colback=yellow!10!white,colframe=red!50!black,colbacktitle=red!40!white,
coltitle=black,center title}}

\newcommand{\leave}[1]{}

\usepackage{mathtools}
\usepackage{relsize}

\newcommand{\toylanguage}{\textbf{langFunny}}
\newcommand{\doublyApply}{\textbf{doublyApply}}
\newcommand{\addToPairAsList}{\textbf{addToPairAsList}}

\newcommand{\ninference}[3]{\inferrule[(#1)]{#2}{#3}}

\newcommand{\ba}{\begin{array}}
\newcommand{\ea}{\end{array}}

\newenvironment{syntax}{\[\ba{l@{\;\;}lcl}}{\ea\]}

\newcommand{\dotspace}{.\,}

\definecolor{ShadowColor}{RGB}{30,150,190}

\makeatletter
\newcommand\Cshadowbox{\VerbBox\@Cshadowbox}
\def\@Cshadowbox#1{%
  \setbox\@fancybox\hbox{\fbox{#1}}%
  \leavevmode\vbox{%
    \offinterlineskip
    \dimen@=\shadowsize
    \advance\dimen@ .5\fboxrule
    \hbox{\copy\@fancybox\kern.5\fboxrule\lower\shadowsize\hbox{%
      \color{ShadowColor}\vrule \@height\ht\@fancybox \@depth\dp\@fancybox \@width\dimen@}}%
    \vskip\dimexpr-\dimen@+0.5\fboxrule\relax
    \moveright\shadowsize\vbox{%
      \color{ShadowColor}\hrule \@width\wd\@fancybox \@height\dimen@}}}
\makeatother

\newcommand{\typeOf}{\vdash}
\newcommand{\step}{\longrightarrow}

\newcommand{\type}{\vdash\,}

\definecolor{lightblue}{rgb}{0.25,0.25,1}

\definecolor{lightgray}{gray}{0.9}
\definecolor{darkergrey}{rgb}{0.75, 0.75, 0.75}
\newcommand{\HI}[1]{\colorbox{darkergrey}{#1}}

\newcommand{\key}[1]{\ensuremath{\mathtt{#1}}}

\newcommand{\Bool}{\key{Bool}}

\newcommand{\lam}[1]{\lambda #1 \dotspace}
\newcommand{\app}{\;}
\newcommand{\of}{{:}}

\definecolor{lightgray}{gray}{0.9}

\newcommand{\Ldl}{\mathcal{L}}

\newcommand{\emptyLdl}[1]{\Ldl^{\epsilon}}

%% file: teaching_paper.bbl
\begin{thebibliography}{10}
\providecommand{\bibitemdeclare}[2]{}
\providecommand{\surnamestart}{}
\providecommand{\surnameend}{}
\providecommand{\urlprefix}{Available at }
\providecommand{\url}[1]{\texttt{#1}}
\providecommand{\href}[2]{\texttt{#2}}
\providecommand{\urlalt}[2]{\href{#1}{#2}}
\providecommand{\doi}[1]{doi:\urlalt{http://dx.doi.org/#1}{#1}}
\providecommand{\bibinfo}[2]{#2}

\bibitemdeclare{book}{Abadi:1996}
\bibitem{Abadi:1996}
\bibinfo{author}{Mart{\'{\i}}n \surnamestart Abadi\surnameend} \&
  \bibinfo{author}{Luca \surnamestart Cardelli\surnameend}
  (\bibinfo{year}{1996}): \emph{\bibinfo{title}{A Theory of Objects}},
  \bibinfo{edition}{2nd} edition.
\newblock \bibinfo{series}{Monographs in Computer Science},
  \bibinfo{publisher}{Springer-Verlag}, \doi{10.1007/978-1-4419-8598-9}.

\bibitemdeclare{article}{Amadio:1993}
\bibitem{Amadio:1993}
\bibinfo{author}{Roberto~M. \surnamestart Amadio\surnameend} \&
  \bibinfo{author}{Luca \surnamestart Cardelli\surnameend}
  (\bibinfo{year}{1993}): \emph{\bibinfo{title}{Subtyping Recursive Types}}.
\newblock {\sl \bibinfo{journal}{ACM Trans. Program. Lang. Syst.}}
  \bibinfo{volume}{15}(\bibinfo{number}{4}), pp. \bibinfo{pages}{575--631},
  \doi{10.1145/155183.155231}.

\bibitemdeclare{inproceedings}{PoulsenM14}
\bibitem{PoulsenM14}
\bibinfo{author}{Casper \surnamestart Bach~Poulsen\surnameend} \&
  \bibinfo{author}{Peter~D. \surnamestart Mosses\surnameend}
  (\bibinfo{year}{2014}): \emph{\bibinfo{title}{Deriving Pretty-Big-Step
  Semantics from Small-Step Semantics}}.
\newblock In: {\sl \bibinfo{booktitle}{Proceedings of the 23rd European
  Symposium on Programming Languages and Systems}}, \bibinfo{volume}{8410},
  \bibinfo{publisher}{Springer-Verlag}, \bibinfo{address}{Berlin, Heidelberg},
  pp. \bibinfo{pages}{270--289}, \doi{10.1007/978-3-642-54833-8_15}.

\bibitemdeclare{inproceedings}{Bracha92}
\bibitem{Bracha92}
\bibinfo{author}{Gilad \surnamestart Bracha\surnameend} \&
  \bibinfo{author}{William \surnamestart Cook\surnameend}
  (\bibinfo{year}{1990}): \emph{\bibinfo{title}{Mixin-Based Inheritance}}.
\newblock In: {\sl \bibinfo{booktitle}{Proceedings of the European Conference
  on Object-Oriented Programming on Object-Oriented Programming Systems,
  Languages, and Applications}}, \bibinfo{series}{OOPSLA/ECOOP '90},
  \bibinfo{publisher}{Association for Computing Machinery},
  \bibinfo{address}{New York, NY, USA}, pp. \bibinfo{pages}{303--311},
  \doi{10.1145/97945.97982}.

\bibitemdeclare{article}{Cardelli84}
\bibitem{Cardelli84}
\bibinfo{author}{Luca \surnamestart Cardelli\surnameend}
  (\bibinfo{year}{1988}): \emph{\bibinfo{title}{A Semantics of Multiple
  Inheritance}}.
\newblock {\sl \bibinfo{journal}{Information and Computation}}
  \bibinfo{volume}{76}(\bibinfo{number}{2/3}), pp. \bibinfo{pages}{138--164},
  \doi{10.1016/0890-5401(88)90007-7}.

\bibitemdeclare{article}{CaMiMaSc1994}
\bibitem{CaMiMaSc1994}
\bibinfo{author}{Luca \surnamestart Cardelli\surnameend},
  \bibinfo{author}{John~C. \surnamestart Mitchell\surnameend},
  \bibinfo{author}{Simone \surnamestart Martini\surnameend} \&
  \bibinfo{author}{Andre \surnamestart Scedrov\surnameend}
  (\bibinfo{year}{1994}): \emph{\bibinfo{title}{An Extension of System F with
  Subtyping}}.
\newblock {\sl \bibinfo{journal}{Information and Computation}}
  \bibinfo{volume}{109}(\bibinfo{number}{1/2}), pp. \bibinfo{pages}{4--56},
  \doi{10.1006/inco.1994.1013}.

\bibitemdeclare{inproceedings}{Cimini:2016aa}
\bibitem{Cimini:2016aa}
\bibinfo{author}{Matteo \surnamestart Cimini\surnameend} \&
  \bibinfo{author}{Jeremy~G. \surnamestart Siek\surnameend}
  (\bibinfo{year}{2016}): \emph{\bibinfo{title}{The Gradualizer: A Methodology
  and Algorithm for Generating Gradual Type Systems}}.
\newblock In: {\sl \bibinfo{booktitle}{Proceedings of the 43rd Annual ACM
  SIGPLAN-SIGACT Symposium on Principles of Programming Languages}},
  \bibinfo{series}{POPL '16}, \bibinfo{publisher}{Association for Computing
  Machinery}, \bibinfo{address}{New York, NY, USA}, pp.
  \bibinfo{pages}{443--455}, \doi{10.1145/2837614.2837632}.

\bibitemdeclare{inproceedings}{Cimini:2017}
\bibitem{Cimini:2017}
\bibinfo{author}{Matteo \surnamestart Cimini\surnameend} \&
  \bibinfo{author}{Jeremy~G. \surnamestart Siek\surnameend}
  (\bibinfo{year}{2017}): \emph{\bibinfo{title}{Automatically Generating the
  Dynamic Semantics of Gradually Typed Languages}}.
\newblock In: {\sl \bibinfo{booktitle}{Proceedings of the 44th ACM SIGPLAN
  Symposium on Principles of Programming Languages}}, \bibinfo{series}{POPL
  2017}, \bibinfo{publisher}{ACM}, \bibinfo{address}{New York, NY, USA}, pp.
  \bibinfo{pages}{789--803}, \doi{10.1145/3093333.3009863}.

\bibitemdeclare{inproceedings}{Ciobaca13}
\bibitem{Ciobaca13}
\bibinfo{author}{\surnamestart \c{S}tefan Ciob\^ac\u{a}\surnameend}
  (\bibinfo{year}{2013}): \emph{\bibinfo{title}{From Small-Step Semantics to
  Big-Step Semantics, Automatically}}.
\newblock In: {\sl \bibinfo{booktitle}{Integrated Formal Methods, 10th
  International Conference, {IFM} 2013, Turku, Finland, June 10-14, 2013.
  Proceedings}}, pp. \bibinfo{pages}{347--361},
  \doi{10.1007/978-3-642-38613-8_24}.

\bibitemdeclare{book}{clrs}
\bibitem{clrs}
\bibinfo{author}{Thomas~H. \surnamestart Cormen\surnameend},
  \bibinfo{author}{Charles~E. \surnamestart Leiserson\surnameend},
  \bibinfo{author}{Ronald~L. \surnamestart Rivest\surnameend} \&
  \bibinfo{author}{Clifford \surnamestart Stein\surnameend}
  (\bibinfo{year}{2009}): \emph{\bibinfo{title}{Introduction to Algorithms}},
  \bibinfo{edition}{3rd} edition.
\newblock \bibinfo{publisher}{The MIT Press}.

\bibitemdeclare{article}{danvy:reduction-free}
\bibitem{danvy:reduction-free}
\bibinfo{author}{Olivier \surnamestart Danvy\surnameend}
  (\bibinfo{year}{2005}): \emph{\bibinfo{title}{From Reduction-based to
  Reduction-free Normalization}}.
\newblock {\sl \bibinfo{journal}{Electronic Notes in Theoretical Computer
  Science}} \bibinfo{volume}{124}(\bibinfo{number}{2}), pp.
  \bibinfo{pages}{79--100}, \doi{10.1016/j.entcs.2005.01.007}.

\bibitemdeclare{inproceedings}{Danvy:2008}
\bibitem{Danvy:2008}
\bibinfo{author}{Olivier \surnamestart Danvy\surnameend}
  (\bibinfo{year}{2008}): \emph{\bibinfo{title}{Defunctionalized Interpreters
  for Programming Languages}}.
\newblock In: {\sl \bibinfo{booktitle}{Proceedings of the 13th ACM SIGPLAN
  International Conference on Functional Programming}}, \bibinfo{series}{ICFP
  '08}, \bibinfo{publisher}{ACM}, \bibinfo{address}{New York, NY, USA}, pp.
  \bibinfo{pages}{131--142}, \doi{10.1145/1411204.1411206}.

\bibitemdeclare{article}{danvy2004refocusing}
\bibitem{danvy2004refocusing}
\bibinfo{author}{Olivier \surnamestart Danvy\surnameend} \&
  \bibinfo{author}{Lasse~R. \surnamestart Nielsen\surnameend}
  (\bibinfo{year}{2004}): \emph{\bibinfo{title}{Refocusing in Reduction
  Semantics}}.
\newblock {\sl \bibinfo{journal}{BRICS Report Series}}
  \bibinfo{volume}{11}(\bibinfo{number}{26}), \doi{10.7146/brics.v11i26.21851}.

\bibitemdeclare{misc}{FelleisenFlatt02}
\bibitem{FelleisenFlatt02}
\bibinfo{author}{Matthias \surnamestart Felleisen\surnameend} \&
  \bibinfo{author}{Matthew \surnamestart Flatt\surnameend}
  (\bibinfo{year}{2006}): \emph{\bibinfo{title}{Programming Languages and
  Lambda Calculi}}.
\newblock \bibinfo{note}{Notes available at
  {\url{https://www.cs.utah.edu/~mflatt/past-courses/cs7520/public_html/s06/notes.pdf}}
  and last accessed in August 2021}.

\bibitemdeclare{article}{krivine2007call}
\bibitem{krivine2007call}
\bibinfo{author}{Jean-Louis \surnamestart Krivine\surnameend}
  (\bibinfo{year}{2007}): \emph{\bibinfo{title}{A Call-by-Name Lambda-Calculus
  Machine}}.
\newblock {\sl \bibinfo{journal}{Higher-Order and Symbolic Computation}}
  \bibinfo{volume}{20}(\bibinfo{number}{3}), pp. \bibinfo{pages}{199--207},
  \doi{10.1007/s10990-007-9018-9}.

\bibitemdeclare{article}{Landin65}
\bibitem{Landin65}
\bibinfo{author}{Peter~J. \surnamestart Landin\surnameend}
  (\bibinfo{year}{1965}): \emph{\bibinfo{title}{Correspondence Between ALGOL 60
  and Church's Lambda-Notation: {Part I}}}.
\newblock {\sl \bibinfo{journal}{Communications of the ACM}}
  \bibinfo{volume}{8}, pp. \bibinfo{pages}{89--101},
  \doi{10.1145/363744.363749}.

\bibitemdeclare{misc}{lnc}
\bibitem{lnc}
\bibinfo{author}{Benjamin \surnamestart Mourad\surnameend}
  (\bibinfo{year}{2019}): \emph{\bibinfo{title}{{Lang-n-Change Tool.}}}
\newblock
  \bibinfo{howpublished}{\url{https://github.com/bmourad01/lang-n-change}}.

\bibitemdeclare{inproceedings}{MouradC20}
\bibitem{MouradC20}
\bibinfo{author}{Benjamin \surnamestart Mourad\surnameend} \&
  \bibinfo{author}{Matteo \surnamestart Cimini\surnameend}
  (\bibinfo{year}{2020}): \emph{\bibinfo{title}{A Calculus for Language
  Transformations}}.
\newblock In: {\sl \bibinfo{booktitle}{46th International Conference on Current
  Trends in Theory and Practice of Informatics ({SOFSEM} 2020)}},
  \bibinfo{publisher}{Springer}, pp. \bibinfo{pages}{547--555},
  \doi{10.1007/978-3-030-38919-2\_44}.

\bibitemdeclare{book}{tapl}
\bibitem{tapl}
\bibinfo{author}{Benjamin~C. \surnamestart Pierce\surnameend}
  (\bibinfo{year}{2002}): \emph{\bibinfo{title}{Types and Programming
  Languages}}, \bibinfo{edition}{1st} edition.
\newblock \bibinfo{publisher}{The MIT Press}.

\bibitemdeclare{inproceedings}{Sewell:2007}
\bibitem{Sewell:2007}
\bibinfo{author}{Peter \surnamestart Sewell\surnameend},
  \bibinfo{author}{Francesco~Zappa \surnamestart Nardelli\surnameend},
  \bibinfo{author}{Scott \surnamestart Owens\surnameend},
  \bibinfo{author}{Gilles \surnamestart Peskine\surnameend},
  \bibinfo{author}{Thomas \surnamestart Ridge\surnameend},
  \bibinfo{author}{Susmit \surnamestart Sarkar\surnameend} \&
  \bibinfo{author}{Rok \surnamestart Strni\v{s}a\surnameend}
  (\bibinfo{year}{2007}): \emph{\bibinfo{title}{Ott: Effective Tool Support for
  the Working Semanticist}}.
\newblock In: {\sl \bibinfo{booktitle}{Proceedings of the 12th ACM SIGPLAN
  International Conference on Functional Programming}}, \bibinfo{series}{ICFP
  '07}, \bibinfo{publisher}{ACM}, \bibinfo{address}{New York, NY, USA}, pp.
  \bibinfo{pages}{1--12}, \doi{10.1145/1291151.1291155}.

\end{thebibliography}
